\documentclass[prb,twocolumn,showpacs,superscriptaddress]{revtex4}

\usepackage{amsmath,amssymb,graphicx}

\begin{document}

\title{Enhancement of mobilities in a pinned multidomain crystal}
\date{\today}
\author{Gwennou Coupier}\altaffiliation{Present address : Laboratoire de Spectrom\'etrie Physique, CNRS \& Universit\'e Joseph Fourier, Grenoble - B.P. 87, 38402 Saint-Martin d'H\`{e}res, France}
\author{Michel Saint Jean}\email{michel.saintjean@paris7.jussieu.fr}
\author{Claudine Guthmann} \affiliation{Laboratoire
Mati\`ere et Syst\`emes Complexes, CNRS \& Universit\'e Paris~7 -
140 rue de Lourmel, 75015 Paris, France}

\begin{abstract}

Mobility properties inside and around degenerate domains of an
elastic lattice partially pinned on a square array of traps are
explored by means of a fully controllable model system of
macroscopic particles. We focus on the different configurations
obtained for filling ratios equal to 1 or 2 when the pinning
strength is lowered. These theoretically expected but never observed
configurations are degenerated, which implies the existence of a
multidomain crystal. We show that the distinction between trapped
and untrapped particles that is made in the case of strong pinning
is not relevant for such a weaker pinning. Indeed, one ought to
distinguish between particles inside or around the domains
associated to positional degeneracies. The possible consequences on
the depinning dynamics of the lattice are discussed.
\end{abstract}

\pacs{74.25.Qt} \maketitle

\section{Introduction}

When submitted to an external force, an elastic lattice of
interacting particles coupled to a regular array of pinning centers
can move as soon as the applied force exceeds a threshold. The
determination of this threshold and of the behavior of the elastic
lattice near it are important and current questions. This threshold
essentially depends upon two parameters : the local deviations of
the particles with respect to their positions in the unpinned
lattice, and the local energetic landscape around the pinned
equilibrium configuration. Thus, knowing how this deformation will
take place and exploring how the particles locally move around their
pinned positions in equilibrium configurations is a first step
towards the understanding of the lattice depinning process.

In this article, we show how the existence of multidomain
configurations would act on the dynamics of an elastic lattice
pinned by a periodic potential of medium amplitude. In order to
explore this question, we have developed a novel experimental system
allowing the direct observation of the interacting particles and a
well controlled pinning process. It will be also presented.

Many elastic systems submitted to a pinning array have been explored
for some years. The most studied of them are the flux lattices in
superconductors since a good pinning of the vortices increases the
critical current in the material. Recently, the focus on such
questions has grown since progress in lithography techniques on
superconducting thin materials has offered the opportunity to design
arrays of holes or magnetic dots, which act as strong pinning
centers where one vortex or even more can be trapped
;~\cite{harada96,grigorenko01,silevitch01,field02,grigorenko03,karapetrov05}
this procedure is much more efficient and potentially controllable
than the weak pinning resulting from intrinsic defects in the
material.~\cite{resumedesordre} In such a situation, a key parameter
is the ratio $f$ between the vortex density, which is controlled by
the magnetic field intensity, and the pinning centers density.
Indeed, measurements of macroscopic characteristics such as the
voltage-current response or of the vortex lattice melting
temperature have shown unambiguously that they strongly  depend upon
$f$ for a square or triangular pinning array and that the pinning is
more efficient for rational $p/q$ values of $f$, where $p$ and $q$
are small.~\cite{welp05}

The dependence of the depinning threshold upon the microscopic
vortex organization and their dynamics has been proved through
semi-analytic or numerical works that were first devoted to the
static problem of the pinned lattice
configuration.~\cite{reichhardt98_1,reichhardt00,reichhardt01_1,reichhardt01_3,laguna01,pogosov03}
Taking the lowest energy equilibrium configuration as a starting
point, the different phases of a driven moving lattice have been
simulated
;~\cite{mov2D,reichhardt01_1,reichhardt01_2,reichhardt01_3} the
resulting quasi-1D movement has also been specifically
studied.~\cite{mov1D} Let us emphasize at this point that, whereas
in most cases the equilibrium configurations are degenerate, the
possible existence of multidomain lattices is seldom taken into
account in these numerical studies.~\cite{notenum} We shall see that
it could be a problem since particles' behaviors strongly differ
depending on whether they are inside the domains or in the walls
between them.

From the experimental point of view, investigating the microscopic
configurations of a vortex lattice is still a challenging issue. The
best imaging of equilibrium configurations has been obtained with
scanning Hall
probe,~\cite{grigorenko01,silevitch01,field02,grigorenko03} scanning
tunneling~\cite{karapetrov05} or Lorentz~\cite{harada96} microscopy.
In these works, configurations of mainly some hundreds of vortices
are imaged for rational filling ratios $f$ close to 1 ($1/4\le f\le
4$). However, even though great efforts have been made in this way,
a direct imaging of (pinned) vortex lattices suffer from two major
drawbacks. First, it is presently impossible to track the vortices
and follow their dynamics ; furthermore, the nature of the
artificial pinning centers used in the works cited above does not
offer the required experimental facility to control the amplitude
and the shape of the pinning potential.

Therefore, a complementary approach based on similar systems has
  appeared to be necessary. Beyond their intrinsic
interest, colloidal crystals can be seen as an attractive
alternative solution. The first attempts in this way have recently
involved colloidal crystals pinned by arrays of optical traps.
Nevertheless, the observed configurations, which are also
numerically predicted,~\cite{reichhardt02_1,autrescollo} are to date
either quite different than the one observed in
superconductors,~\cite{brunner02} or restricted to system of very
small spatial extent.~\cite{mangold03} Moreover, the role of the
hydrodynamic couplings and of the nature of the pinning strength
are, in these systems, still unclear.

In order to bypass these difficulties, we propose here an
alternative experimental system consisting of a few thousand
millimetric charged particles which can be submitted to a controlled
electrostatic pinning potential and a tunable effective temperature
resulting from a mechanical shacking. After a brief description of
the pinning set-up and the experiments which validate its efficiency
(Sec.~\ref{sec:presentation}), we present in
Sec.~\ref{sec:equilibre} the multidomain equilibrium configurations
and their evolution when the amplitude of the pinning strength
decreases in the case of a square array of traps. We mainly focus on
the $f=1$ and $f=2$ cases. Finally, the thermally activated motion
of the particles in these multidomain systems are described and
discussed in Sec.~\ref{sec:mouvement}.

\section{Experimental set-up} \label{sec:presentation}

\begin{figure}

\resizebox{\columnwidth}{!}{\includegraphics{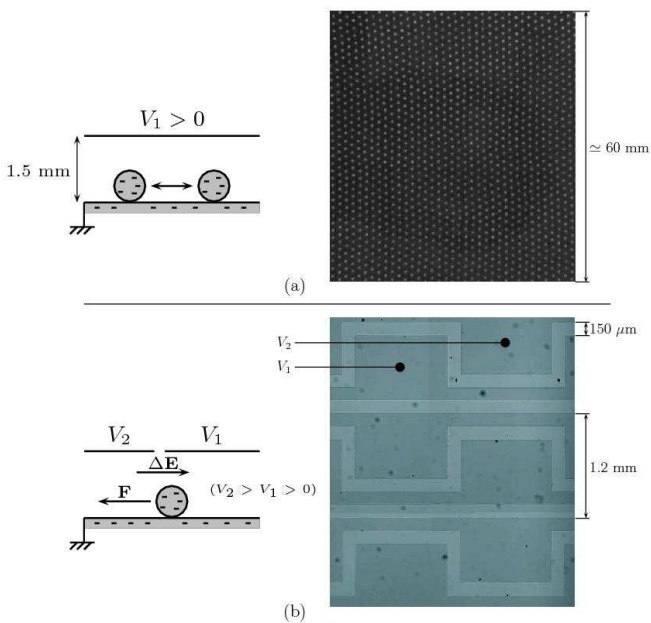}} \caption{Summary
of the two electrostatic forces at stack. (a) : Scheme accounting
for the origin of the repulsive interaction between two balls and
picture of the resulting  triangular lattice. (b) : Scheme
accounting for the apparition of the pinning force under two areas
linked to two different potentials and picture of a few adjacent
patterns in the ITO electrode that is used to study the f = 1 case.}
\label{fig:fig0}
\end{figure}

\begin{figure}
\resizebox{\columnwidth}{!}{\includegraphics{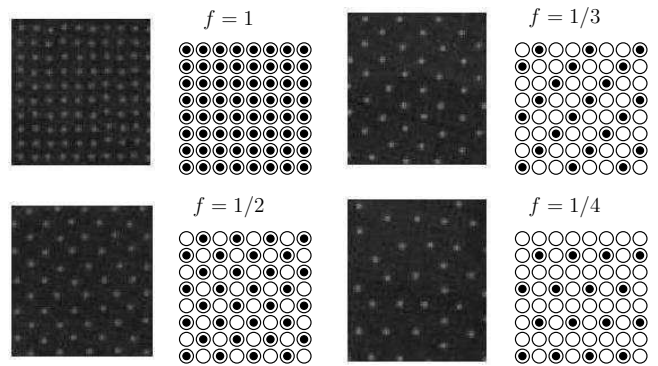}}
\caption{Overview of the equilibrium configurations on a square
array with filling ratios equal to $1,1/2,1/3$ and $1/4$. The
corresponding theoretical schemes for a mono-domain configuration
are also shown. The traps are represented by empty circles, while
the black disks represent the particles.} \label{fig:fig5}
\end{figure}

\begin{figure}
\resizebox{\columnwidth}{!}{\includegraphics{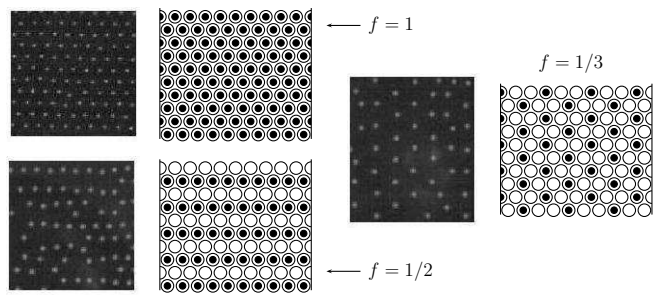}}
\caption{Overview of the equilibrium configurations and
corresponding schemes for a triangular array with filling ratios
equal to $1,1/2$ and $1/3$.} \label{fig:fig6}
\end{figure}

The system  we propose, a so-called "pinned macroscopic Wigner
crystal", is made of a mono-layer of confined charged millimetric
balls. These conducting balls are located on the grounded bottom
electrode of an horizontal plane condenser. When a tunable voltage
$V_1>0$ ($V_1 \simeq 1$ kV) is applied between the top and the
bottom electrodes, the balls get negatively charged and repel each
other (Fig.~\ref{fig:fig0}(a)). In Ref.~\onlinecite{galatola06}, we
have shown that the interball interaction potential $E(r)$ is equal
to $E_0 K_0(r/\lambda)$, where $K_0$ is the modified Bessel function
of the second kind, and $E_0$ is proportional to $V_1^2$. This
interaction between the balls has the same space-dependence as the
interaction between vortices in superconductors, which suggests that
the observed behaviors could be directly mapped onto vortex systems
without the experimental constraints associated to the real ones. To
introduce thermal noise, the whole cell is fixed on loudspeakers
supplied by a white noise voltage. We have thoroughly checked that
the resulting shaking of the balls, due to friction with the bottom
electrode, fulfil the properties one can expect for a thermal
shaking. First, even though all the balls lie on the same solid
substrate, their movement is spatially non correlated, which might
be due to inhomogeneities on the wafer at a microscopic level.
Secondly, the individual trajectory of a single ball which is free
or trapped in a parabolic well can be described through Langevin
formalism.~\cite{coupier06} Stationary states are reached in a few
tenth of second. They are characterized by an effective temperature
directly controlled by the shaking amplitude : the energy
distribution is given by Boltzmann statistics. This was proved on
confined small islands of balls, that can be seen as two-level
systems when considering their two first equilibrium configurations,
characterized by concentric shells of varying number of
balls.~\cite{coupier05} This effective temperature was calibrated
and is measured in-situ.

When submitted to a low effective temperature, the ensemble of balls
can organize and form a defect-free triangular lattice of a few
thousand particles.~\cite{stjean04} By insulating a periodic set of
small patterns in the top electrode we can link it to a different
tunable voltage $V_2>V_1$, and create an array of pinning sites, as
the local potential difference results in an attractive force on a
ball (Fig.~\ref{fig:fig0}(b)). Since the net charge on the ball
around the trap is roughly proportional to the mean potential $V$
between $V_1$ and $V_2$ , the amplitude of the pinning potential can
be considered as proportional to $V\Delta V$, where $\Delta
V=V_2-V_1$ comes from the gradient of the electrostatic potential
between inside and outside the trap. Therefore, the relevant
parameter to discuss the relative strength of the pinning compared
with the elastic stiffness will be $\eta_V=V\Delta V/V^2=\Delta
V/V$, which is fully controlled.

In our experiments, the lattice spacing $a_0$ in the unpinned
lattice (made of nearly 2000 balls) is 1.82 mm, to be compared with
the screening length $\lambda=0.48$ mm. The periodicity of the
square pinning array is then  chosen so that the filling ratio $f$
is equal to 1 or 2. The width of the patterns is such that the range
of the pinning potential associated to one site is of the same order
as the interball distance.

In order to test first the efficiency of this pinning device we have
observed the equilibrium configurations obtained when starting from
$f=1$ and taking off some balls so that $f\le 1$, with high values
for the relative pinning strength ($\eta_V \ge 0.25$). These
configurations have been compared to those reported in the
literature.

Our experimental equilibrium configurations are shown for some
fractional filling ratios on Fig.~\ref{fig:fig5} for a square
pinning array and on Fig.~\ref{fig:fig6} for a triangular array. In
the first case, they are similar to those observed in
superconductors~\cite{harada96,field02,grigorenko03} and numerically
predicted.~\cite{reichhardt01_1,pogosov03} In the second case, they
are also similar to those numerically
predicted.~\cite{reichhardt01_1} These results confirm the ability
of our system to be generic of elastic systems and to describe well
the behavior of flux lattices in superconductors.

\section{Multidomain equilibrium configurations }\label{sec:equilibre}

The existence of multidomain equilibrium configurations depends upon
different parameters.

\begin{figure}
\resizebox{0.95\columnwidth}{!}{\includegraphics{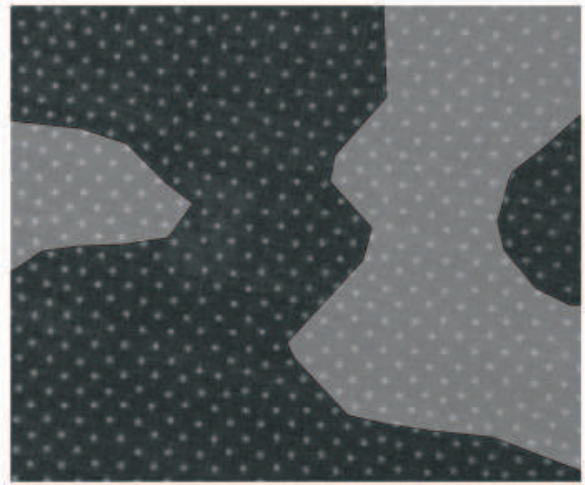}}
\caption{Snapshot of the crystal in a quasi-equilibrium state with
$f=1/2$ and a square pinning array. The theoretical configuration is
2-fold degenerate, which leads to the existence of two kinds of
domains (black or gray background colors).} \label{fig:fig7}
\end{figure}

For $f<1$, the equilibrium configurations are structurally
degenerate since the organized pinned lattice can be pinned on
various subarrays of traps whatever the shape and the intensity of
the pinning potential. Therefore it is experimentally difficult to
obtained large coherent domains and multidomain systems are often
observed as for instance in our system (Fig.~\ref{fig:fig7}), or in
superconductors, where a multidomain experimental configuration of a
$f=1/2$ crystal of more than $10^4$ vortices has clearly been
imaged.\cite{field02} On the contrary, according to what stated in
Ref.~\onlinecite{reichhardt01_1}, it seems to be difficult to find
these multidomain configurations through numerical simulations.

\begin{figure*}
\resizebox{1.9\columnwidth}{!}{\includegraphics{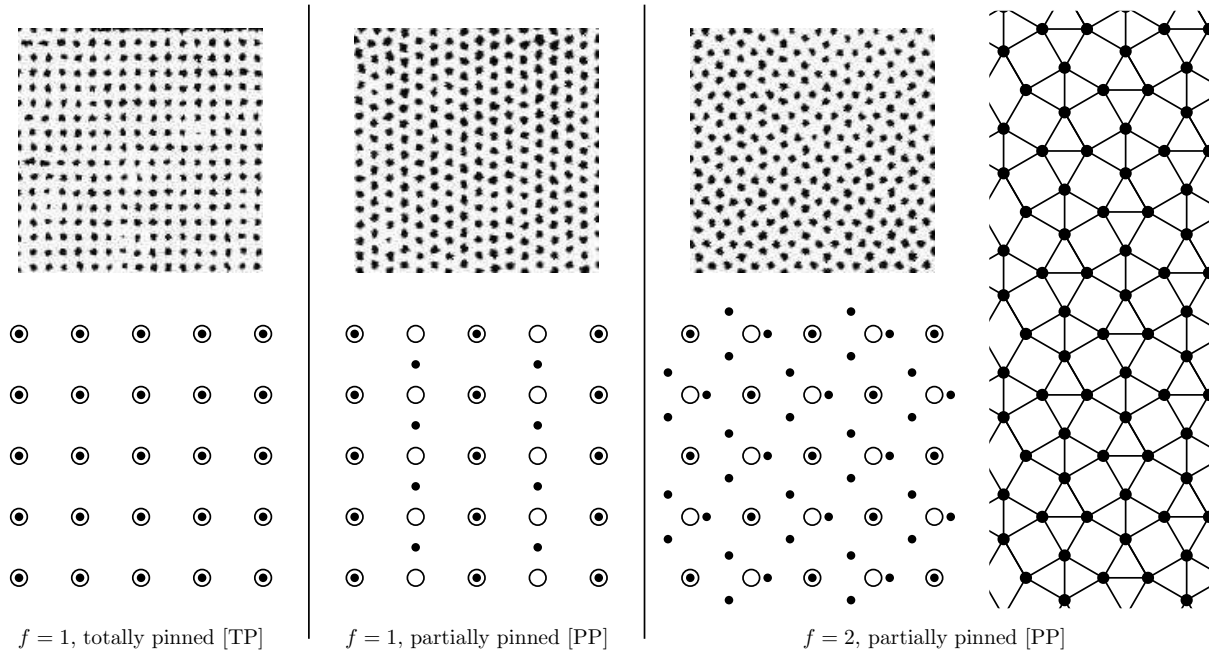}}
\caption{Thermally activated trajectories around equilibrium
configurations and corresponding schemes. For $f = 1$, the
transition between both configurations is obtained by tuning
$\eta_V$. The Delaunay triangulation of the partially pinned
configuration for $f = 2$ is also shown.} \label{fig:fig1}
\end{figure*}

Beyond these easily predictable cases, theoretical studies suggest
that for $f\ge 1$, lowering the amplitude of the pinning strength or
widening the traps can lead to partially pinned (PP) configurations,
where only a fraction of the vortices are located in the middle of
the pinning sites,~\cite{pogosov03} allowing multidomain
configurations. Such configurations have never been observed in
superconductors, perhaps because the pinning intensity is too high ;
however our experiments in which the pinning strength can be tuned
confirm this assumption.

\begin{figure}
\resizebox{\columnwidth}{!}{\includegraphics{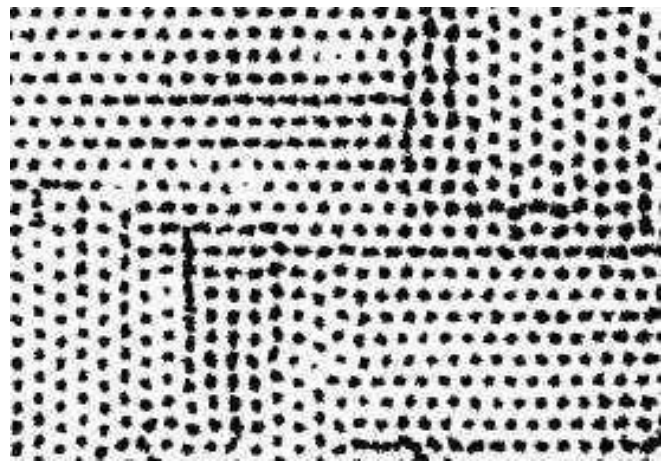}}
\caption{Thermally activated trajectories in a multidomain partially
pinned configuration for $f = 1$. } \label{fig:fig2}
\end{figure}

For $f =1$, a sharp transition between the totally pinned (TP) and
the partially pinned configuration is observed when lowering
$\eta_V$ ($\eta<0.15$). The latter configuration consists in an
alternation of rows of trapped or untrapped balls, so that the
triangular symmetry of the unpinned lattice is almost restored, the
position of the untrapped particle between two free traps slightly
depending  on the relative pinning strength (Fig.~\ref{fig:fig1}).
Let us now emphasize that this partially pinned configurations is
degenerate since there are two possibilities for the orientation of
the unpinned rows (Fig.~\ref{fig:fig2}).

Note that we don't discuss here the existence of a "floating
crystal", a triangular lattice slightly distorted by the traps : in
the literature, its orientation is controversial, as it can be
aligned with the axes of the underlying pinning
array,~\cite{reichhardt01_2} or slightly rotated.~\cite{pogosov03}
Experimentally, this orientation should depend strongly on the
boundary conditions, unlike the pinned configurations. Anyhow, as
for $f=1$,~\cite{pogosov03} one should expect this configuration to
be favored when the traps widen and can include many particles :
then, a triangular-like structure will be obviously preferred.

In the $f = 2$ case, a "totally pinned" configuration (TP), where
all the traps are occupied and the remaining particles are located
in the center regions between the traps is usually observed in
superconductors,~\cite{harada96,grigorenko01} and also numerically
predicted.~\cite{reichhardt98_1} We have also obtained this
configuration with a strong pinning ($\eta\ge 4$). For traps with
larger extent totally pinned configurations switch to a regular
array of dipoles centered on the traps as it was shown by numerical
simulations for vortices with logarithmic
interaction.~\cite{reichhardt00} This configuration could be the one
observed by Field et al., where two flux quanta are trapped in each
pinning center.~\cite{field02} Similar configurations seem to be
also observed in colloidal systems. Indeed, the latter affirmation
is an extrapolation of the observations made in
Ref.~\onlinecite{brunner02}, where the $f=3$ case is studied and an
array of trimers is observed. This extrapolation is confirmed by
numerical simulations made by Reichhardt et
al.~\cite{reichhardt02_1} In these cases, the degeneracy is linked
to the rotation of the dimers, while the TP configuration is not
degenerate. These results indicate that the existence of multidomain
systems depends upon the intensity and the shape of the pinning
potential.

Since in our experiment we can tune the pinning intensity (the trap
shape being fixed) we have explored the evolution of the TP
configuration as this intensity decreases. For not too high values
of $\eta_V$, we have  observed an unexpected and interesting
partially pinned (PP) configuration shown on Fig.~\ref{fig:fig1}.
Only half the traps are occupied while the other ones are the
centers of trimers, effectively trapping three particles instead of
one. It results from this an elegant four-fold configuration in
which each particles has got exactly five equidistant neighbors.
These patterns result from a compromise between the symmetry imposed
by the pinning array and the symmetry of the unpinned lattice, which
is six-fold. Since there are two possibilities for the choice of the
subarray of occupied traps and, for each selected subarray, four
possibilities for the orientation of the trimers, the system is
highly degenerated and, as a consequence, a multidomain
configuration will appear in the pinned lattice.\\
\begin{figure}
\resizebox{\columnwidth}{!}{\includegraphics{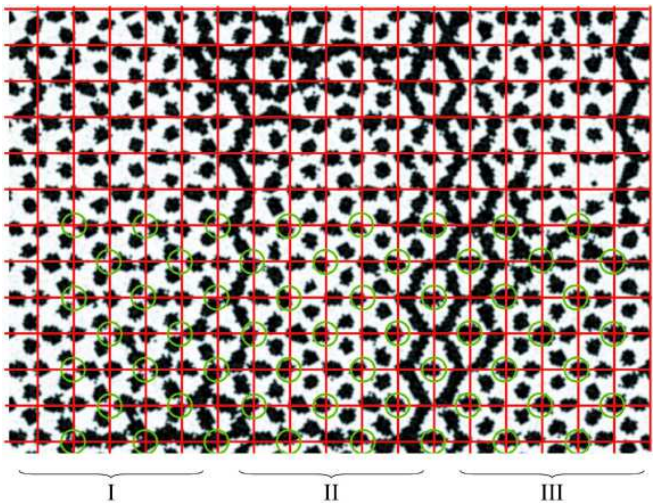}} \caption{(color
online)Thermally activated trajectories in a multidomain partially
pinned configuration for $f = 2$. The traps are located at the
intersections of the red lines. The green circles mark one of the
two subarrays that can be occupied by particles.} \label{fig:fig3}
\end{figure}

The multidomain partially pinned configurations are observed in our
whole system as shown on Fig.~\ref{fig:fig2} for $f = 1$ and
Fig.~\ref{fig:fig3} for $f=2$, where three domains are present. In
the latter, the difference between domains I and II lies in the
choice of the occupied subarray. Between domains II and III, the
trimers do not have the same orientation. In both cases, the walls
are marked by features looking like "eyes" constituted by a well
trapped particle surrounded by very mobile ones. We shall return to
this last point in the following.

In these configurations, the domains' sizes increase for high $V$,
thus for high intensities of the forces at stack. Notice furthermore
that these configurations are highly symmetric, the walls between
the domains being mainly orientated along the principle axes of the
pinning lattice. These equilibrium multidomain configurations are
very stable and can be considered as quasi-static because of the
presence of pinned particles. The evolution in shape and size of
each domain is much slower than in the case of a disoriented cluster
in an unpinned lattice that was discussed in
Ref.~\onlinecite{stjean04}.

In order to determine the conditions required to observe such
configurations, we have calculated the phase diagrams of the pinned
macroscopic Wigner  lattices, the two relevant parameter being the
relative pinning intensity $E_p/E_0$ and the width $\sigma$ of each
electrostatic trap.

Since the analytical determination of these attracting potentials is
a tricky problem, we have calculated the phase diagram for different
profile and compared them to the experimental results.

This procedure is relevant to find the depth $E_p$ and width
$\sigma$ of the traps since the obtained phase diagrams strongly
differ according to the choice in the profile used for its
calculation as it was shown in Ref.~\onlinecite{pogosov03} for $f=1$

Various shapes have been proposed in the literature : in the
numerical simulations of flux lattices this potential is either a
truncated parabola or a
gaussian,~\cite{reichhardt98_1,reichhardt00,reichhardt01_1,pogosov03}
whereas the optical traps in colloidal systems seem to be well
described by sine functions.~\cite{reichhardt02_1} Our calculations
show that a good agreement is obtained for traps with gaussian
profiles ${E}_g (r)=-E_p \exp\big[-r^2/(2\sigma^2)\big]$.

\begin{figure}
\resizebox{\columnwidth}{!}{\includegraphics{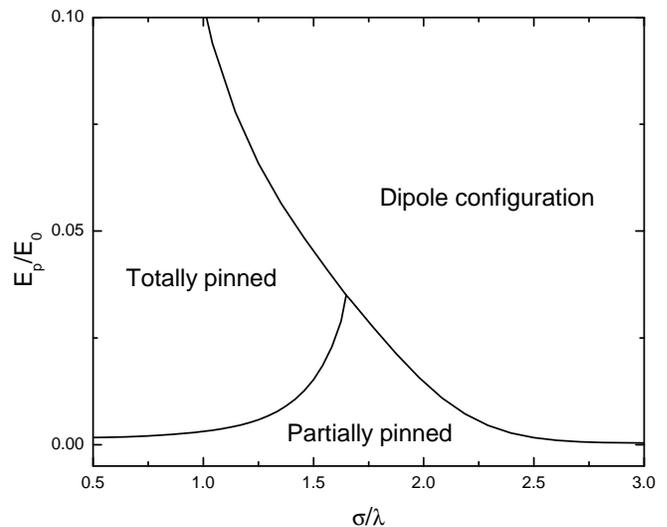}} \caption{Phase
diagram for the three pinned configurations that are found here and
in the literature for $f=2$ and a square pinning array. The values
for $a_0$ and $\lambda$ are the experimental ones. The profile of
the wells is gaussian, with a depth $E_p$ and a width $\sigma$.}
\label{fig:fig8}
\end{figure}

In particular, using this trap profile, we have calculated the phase
diagram for $f=2$ for which we consider three possible
configurations : the totally pinned configuration, the partially
pinned one and the dipole configuration found in
Ref.~\onlinecite{reichhardt00}. The resulting phase diagram for a
gaussian profile is shown on Fig.~\ref{fig:fig8}.

It can be seen from the diagram that the partially pinned
configuration is favored relatively to the totally pinned one for
sufficiently wide or shallow traps. Perhaps the dipole configuration
could not be observed in our system because the traps are not wide
enough, or because $\eta_V$ has a upper bound by construction.

Let us nevertheless indicate that such a phase diagram has to be
considered more as a guide for the experiment than a tool for
precise predictions : indeed these phase diagrams are sensitive to
the details of the unknown profile of the traps and one cannot
assert that the real traps in our experiments or in flux lattices
can be described by the simple functions that are usually tried.

\begin{figure}
\resizebox{\columnwidth}{!}{\includegraphics{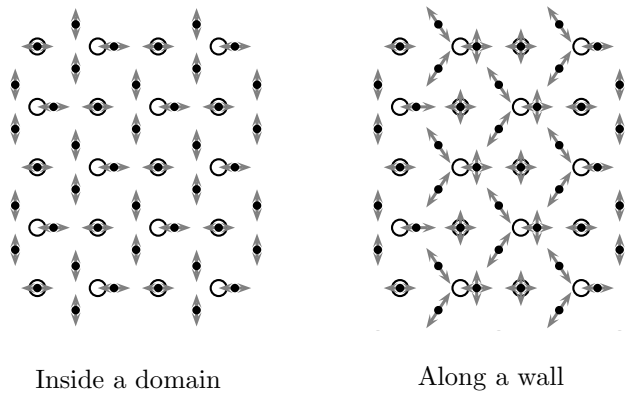}} \caption{Schemes
of the main movement of the particles inside a domain or along a
domain wall at low temperature. Note that the lacking or excess
particles of the wall are not represented.} \label{fig:fig9}
\end{figure}

\begin{figure}
\resizebox{\columnwidth}{!}{\includegraphics{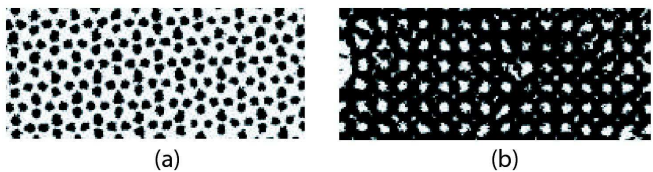}}
\caption{Trajectories inside an homogeneous domain of the partially
pinned configuration for $f = 2$. In (b) the temperature is twice
the temperature in (a).} \label{fig:fig4}
\end{figure}

\section{Particles dynamics between and along the domains
walls}\label{sec:mouvement}

The presence of quasi-static domains leads to the existence of two
kinds of particles, viz. those inside the domains, and those near
the walls. As discussed by Grigorenko et al.,\cite{grigorenko03}
lacking or excess particles are required for the adjustment of the
domains and the construction of the walls. Thus, these two kinds of
particles are associated to different local densities. Our
experiments show that even if these density variations are tiny,
they induce important differences in the dynamics of the balls, even
in the absence of an external force. More precisely, the movements
in or alongside the domains strongly differ from the view of the
associated mobility and the direction of the movement.

In the $f=2$ case (Fig.~\ref{fig:fig3}), if one first looks inside
the domains, the trapped or untrapped balls (viz. those belonging to
a trimer) are stable and their dynamics essentially result from the
thermal activation. As it can be seen on the trajectory spots, the
amplitude of the movements of the trapped or of the trimer particles
are equal. We can conclude from this result that the particle status
towards the traps is not relevant to discuss its dynamical
properties. Underline that this behavior can be associated to the
conditions under which the partially pinned configuration is
energetically favorable, notably the necessity to have wide traps.
In already studied totally pinned
configurations,~\cite{reichhardt01_2,mangold03} interstitial
particles are much more mobile than trapped ones.

The dynamics along the walls is pretty different. Indeed, the "eyes"
in the walls are constituted by a well trapped particle, whose
wandering amplitude is similar to the one inside a domain,
surrounded by much more mobile untrapped particles. This high
mobility can be associated to the local metastability linked with a
small density variation. This makes the fact that the trapped
particles remain in their place and are not chased by their
neighbors all the more surprising.

Furthermore, the favored directions are also pretty different
according to the class of the particles. These directions are
schematized on Fig.~\ref{fig:fig9}. Inside a domain, all the
 movements have a privileged direction following the
symmetry of the underlying pinning array. Notice that for the
untrapped balls, this movement is surprising since it is not
orientated towards the traps. This appears clearly when the
temperature increases, the symmetry of the underlying pinning array
being then directly recovered through the trajectories mapping
(Fig.~\ref{fig:fig4}). Note that this is completely different than
the "two-stage melting" that is found in the dipole
configuration.~\cite{reichhardt00} On the other hand, the untrapped
particles near a metastable area are clearly following trajectories
linking two adjacent empty pinning centers. We think that this
phenomenon should be directly linked with the existence of trimers
which are in a very uncomfortable position and are highly perturbed
by a density variation. By contrast, for $f = 1$, the movements
along the walls are also enhanced relatively to the movements inside
the domains but are still respecting the same 4-fold underlying
symmetry as inside the domains (Fig.~\ref{fig:fig2}).

As a conclusion, this study performed with a macroscopic Wigner
crystal has shown without ambiguity that the existence of domains
must be taken into account to discuss the particles dynamics in a
pinned elastic lattice. Those multidomain configurations appear as
soon as the pinning amplitude diminishes and implies degenerate
equilibrium configurations. The tracking of the particles has shown
that the discrimination that should be made between trapped and
interstitial particles in the totally pinned case is not relevant in
a weaker pinning case for which one should discuss the dynamics by
discriminating between particles in stable or metastable situations,
respectively inside or at the boundaries of the domains. This
suggests that the existence of domains must be taken into account in
further experimental or numerical studies of the moving phases of a
partially pinned lattice submitted to a depinning force.
Consequences could be first an easier depinning along the walls,
which would then induce anisotropy since the orientations of the
walls are  given by the principle axes of the pinning array. In
return, depinning of the lattice could remove the degeneracies,
which would then imply isotropy. In
Ref.~\onlinecite{reichhardt01_1}, the nature of the flow is
discussed according to the values of the filling ratio $f$. It is
shown that if $f$ is rational, this flow has elastic properties,
whereas plasticity is invoked when $f$ is irrational. Even though
the "macroscopic filling ratio" is rational here, the local ratio
near the boundaries is irrational and thus, the nature of the
depinning dynamics could become very complex with a coexistence of
elastic and plastic flows in the same lattice.


\begin{thebibliography}{46}
\bibitem{notenum}This might be due to the limited number of particles that
simulations are able to deal with up to date.
\end{thebibliography}
\end{document}